\documentclass[twocolumn,showpacs,aps,prl,superscriptaddress]{revtex4}
\usepackage{graphicx}
\usepackage{dcolumn}
\usepackage{amsmath}
\usepackage{supertabular}
\usepackage{multirow}
\usepackage{multicols}

\begin{document}

\affiliation{Argonne National Laboratory, Argonne, Illinois 60439}%
\affiliation{Brookhaven National Laboratory, Upton, New York 11973}%
\affiliation{University of Birmingham, Birmingham, United Kingdom}%
\affiliation{University of California, Berkeley, California 94720}%
\affiliation{University of California, Davis, California 95616}%
\affiliation{University of California, Los Angeles, California 90095}%
\affiliation{Carnegie Mellon University, Pittsburgh, Pennsylvania 15213}%
\affiliation{Creighton University, Omaha, Nebraska 68178}%
\affiliation{Nuclear Physics Institute AS CR, \v{R}e\v{z}/Prague, Czech Republic}%
\affiliation{Laboratory for High Energy (JINR), Dubna, Russia}%
\affiliation{Particle Physics Laboratory (JINR), Dubna, Russia}%
\affiliation{University of Frankfurt, Frankfurt, Germany}%
\affiliation{Indiana University, Bloomington, Indiana 47408}%
\affiliation{Insitute  of Physics, Bhubaneswar 751005, India}%
\affiliation{Institut de Recherches Subatomiques, Strasbourg, France}%
\affiliation{University of Jammu, Jammu 180001, India}%
\affiliation{Kent State University, Kent, Ohio 44242}%
\affiliation{Lawrence Berkeley National Laboratory, Berkeley, California 94720}%
\affiliation{Max-Planck-Institut f\"ur Physik, Munich, Germany}%
\affiliation{Michigan State University, East Lansing, Michigan 48824}%
\affiliation{Moscow Engineering Physics Institute, Moscow Russia}%
\affiliation{City College of New York, New York City, New York 10031}%
\affiliation{NIKHEF, Amsterdam, The Netherlands}%
\affiliation{Ohio State University, Columbus, Ohio 43210}%
\affiliation{Panjab University, Chandigarh 160014, India}%
\affiliation{Pennsylvania State University, University Park, Pennsylvania 16802}%
\affiliation{Institute of High Energy Physics, Protvino, Russia}%
\affiliation{Purdue University, West Lafayette, Indiana 47907}%
\affiliation{University of Rajasthan, Jaipur 302004, India}%
\affiliation{Rice University, Houston, Texas 77251}%
\affiliation{Universidade de Sao Paulo, Sao Paulo, Brazil}%
\affiliation{University of Science \& Technology of China, Anhui 230027, China}%
\affiliation{Shanghai Institute of Nuclear Research, Shanghai 201800, China}%
\affiliation{SUBATECH, Nantes, France}%
\affiliation{Texas A\&M, College Station, Texas 77843}%
\affiliation{University of Texas, Austin, Texas 78712}%
\affiliation{Valparaiso University, Valparaiso, Indiana 46383}%
\affiliation{Variable Energy Cyclotron Centre, Kolkata 700064, India}%
\affiliation{Warsaw University of Technology, Warsaw, Poland}%
\affiliation{University of Washington, Seattle, Washington 98195}%
\affiliation{Wayne State University, Detroit, Michigan 48201}%
\affiliation{Institute of Particle Physics, CCNU (HZNU), Wuhan, 430079 China}%
\affiliation{Yale University, New Haven, Connecticut 06520}%
\affiliation{University of Zagreb, Zagreb, HR-10002, Croatia}%

\author{J.~Adams}\affiliation{University of Birmingham, Birmingham, United Kingdom}%
\author{C.~Adler}\affiliation{University of Frankfurt, Frankfurt, Germany}%
\author{M.M.~Aggarwal}\affiliation{Panjab University, Chandigarh 160014, India}%
\author{Z.~Ahammed}\affiliation{Variable Energy Cyclotron Centre, Kolkata 700064, India}%
\author{J.~Amonett}\affiliation{Kent State University, Kent, Ohio 44242}%
\author{B.D.~Anderson}\affiliation{Kent State University, Kent, Ohio 44242}%
\author{M.~Anderson}\affiliation{University of California, Davis, California 95616}%
\author{D.~Arkhipkin}\affiliation{Particle Physics Laboratory (JINR), Dubna, Russia}%
\author{G.S.~Averichev}\affiliation{Laboratory for High Energy (JINR), Dubna, Russia}%
\author{S.K.~Badyal}\affiliation{University of Jammu, Jammu 180001, India}%
% ...
%

\collaboration{STAR Collaboration}\noaffiliation

\title{ \vspace*{-0.5cm}
Particle-type dependence of azimuthal anisotropy and nuclear
modification of particle production in Au+Au collisions at
$\sqrt{s_{_{NN}}} = 200$~GeV }\noaffiliation

\date{\today}

\begin{abstract}

We present STAR measurements of the azimuthal anisotropy parameter
$v_2$ and the binary-collision scaled centrality ratio $R_{CP}$ for
kaons and lambdas ($\Lambda+\overline{\Lambda}$) at mid-rapidity in
Au+Au collisions at $\sqrt{s_{_{NN}}}=200$~GeV.  In combination, the
$v_2$ and $R_{CP}$ particle-type dependencies contradict expectations
from partonic energy loss followed by standard fragmentation in
vacuum.  We establish $p_T \approx 5$~GeV/c as the value where the
centrality dependent baryon enhancement ends.  The $K_S^0$ and
$\Lambda+\overline{\Lambda}$ $v_2$ values are consistent with
expectations of constituent-quark-number scaling from models of hadron
fromation by parton coalescence or recombination.


\end{abstract}

  \maketitle

\vspace{0.5cm}


The azimuthal anisotropy and system-size dependence of identified
particle yields at moderate and high transverse momentum ($p_T$) may
provide insight into the existence and properties of a deconfined
partonic state in ultra-relativistic heavy-ion
collisions~\cite{CoalLinv2,CoalVoloshinv2,CoalMullerRaa,CoalKoRatios}.
The azimuthal anisotropy parameter $v_2$ is thought to be sensitive to
the earliest stages of heavy-ion
collisions~\cite{hydroOllitrault92}. The parameters $v_n$ are derived
from a Fourier expansion of the azimuthal component ($\phi$) of the
momentum-space distribution; $dN/d\phi \propto 1 +
\sideset{}{_n}\sum\nolimits 2v_n\cos n\left (\phi-\Psi_{RP} \right )$,
where $\Psi_{RP}$ is the reaction-plane angle. Previous measurements
at the Relativistic Heavy-Ion Collider (RHIC) established that $v_2$
for charged hadrons rises with $p_T$ for $p_T<2$~GeV/c and then
saturates~\cite{aihong,Chgv2}. At low $p_T$ ($p_T < 1$~GeV/c), the
dependence of $v_2$ on particle mass~\cite{Stv2pid,v0v2130} is
consistent with hydrodynamic calculations where local thermal
equilibrium of partons has been
assumed~\cite{hydroOllitrault92,hydroPasi01,hydroShuryak01}.

Surface emission has been considered in relation to the large
saturated $v_2$ at higher $p_T$~\cite{SurfShuryak}. The existence of a
dense, opaque medium in which fast partons suffer energy loss can
naturally lead to a surface emission pattern.

Parton energy loss in a dense medium may also suppress high $p_T$
particle yields in central Au+Au collisions at RHIC~\cite{dEdx}. High
$p_T$ particles are produced from initial hard parton scatterings
whose cross-sections are assumed to be proportional to the number of
binary nucleon-nucleon collisions $\mathrm{N_{bin}}$. The
$\mathrm{N_{bin}}$ scaled centrality ratio $R_{CP}$ is a measure of
the particle production's dependence on the collision system's size and
density:
\begin{equation*}
R_{CP}(p_T) = 
\frac{\left [\left (dN/dp_T\right )/\mathrm{N_{bin}}\right ]^{Central}}
     {\left [\left (dN/dp_T\right )/\mathrm{N_{bin}}\right ]^{Peripheral}},
\end{equation*}
where $R_{CP} = 1$ if particle production is equivalent to a
superposition of independent nucleon-nucleon collisions.
In central Au+Au collisions at $\sqrt{s_{NN}}=130$ and 200~GeV, the
moderate and high $p_T$ neutral pion and charged hadron yields are
suppressed relative to $\mathrm{N_{bin}}$ scaling (\textit{i.e.}
$R_{CP}$ and the closely related nuclear modification factor $R_{AA}$
are below unity)~\cite{HighPt130,highpt200}.
For $1<p_T<4.5$~GeV/c, the neutral pion yield is more strongly
suppressed than the charged hadron yield, indicating a particle-type
dependence for $R_{CP}$. Within the framework of parton energy loss
followed by standard fragmentation, the suppression and $v_2$ both
reflect the magnitude of the energy loss. The particle-type dependence
of $v_2$ and $R_{CP}$ will provide a stringent test for energy loss
models.

Quark coalescence or
recombination~\cite{CoalVoloshinv2,CoalLinv2,CoalMullerRaa,CoalKoRatios}
models for hadron formation are an alternative to the fragmentation
models commonly used in energy loss calculations~\cite{dEdx}.  In
these models, a particle-type dependence develops at hadronization
with baryons developing a larger $v_2$ and $R_{CP}$ than mesons. In
this letter we present measurements of $v_2$ and $R_{CP}$ at
mid-rapidity ($|y|<1$) for $K_S^0$ and $\Lambda+\overline{\Lambda}$
for $0.2<p_T<6.5$ and $0.4<p_T<6.0$~GeV/c respectively
%
%
along with $R_{CP}$ for $K^{\pm}$ from $0.2<p_T<3.0$~GeV/c in
Au+Au collisions at $\sqrt{s_{_{NN}}}=200$~GeV.
The $K_S^0$ and $\Lambda+\overline{\Lambda}$ analysis extends the
measurement of $v_2$ and $R_{CP}$ for identified particles to a
$p_T$ range where previously only neutral pion $R_{CP}$ had been
measured and establishes the particle-type dependence of $v_2$ and
$R_{CP}$ at intermediate $p_T$ (1.5--4.0~GeV/c) and high $p_T$
($p_T>5$~GeV/c).

This analysis uses $1.6\times 10^{6}$ minimum--bias trigger events
and $1.5\times 10^{6}$ central trigger events from the Solenoidal
Tracker at RHIC (STAR) experiment~\cite{STAR}. The $K_{S}^{0}$ and
$\Lambda (\overline{\Lambda})$ were reconstructed from the
topology of the decay channels, $K_{S}^{0}\rightarrow \pi ^{+}+\pi
^{-}$ and $\Lambda (\overline{\Lambda })\rightarrow p+\pi
^{-}(\overline{p}+\pi ^{+})$.  A detailed description of the
analysis, such as track quality, decay vertex topology cuts, and
detection efficiency, can be found in
Refs.~\cite{v0v2130,LLbar130,sorensen}. The $K^{\pm}$ are
identified from one-prong decays as described in
Ref.~\cite{StarKaon}.
For both $v_2$ and $R_{CP}$, no difference is seen between
$\Lambda$ and $\overline{\Lambda}$ within statistical errors.
The reaction-plane angle is estimated from the azimuthal distribution
of primary tracks~\cite{art} with
$0.1<p_T<2.0$~GeV/c and $|\eta|<1.0$, where $\eta$ is the
pseudorapidity. To avoid autocorrelations, tracks associated with a
$K_S^0$, $\Lambda$ or $\overline{\Lambda}$ decay vertex are excluded
from the calculation of $\Psi_{RP}$.

\begin{table}[hbt]
\caption{The relative systematic errors (\%) from background (bg) and non-flow
effects (n-f) for $v_2$ (0--80\%), and from background and the
efficiency calculation (eff) for $R_{CP}$ (0--5\%/40--60\%) are
listed for three $p_T$ values. } \label{syserr}
\begin{tabular}{l|ccc|cc|ccc}
\toprule
~ & \multicolumn{3}{c}{$K_S^0$} \vline &  \multicolumn{2}{c}{$K^{\pm}$} \vline & \multicolumn{3}{c}{$\Lambda+\overline{\Lambda}$} \\
\colrule
$p_T$ (GeV/c) & 1.0 & 2.5 & 4.0 & 1.0 & 2.5 & 1.0 & 2.5 & 4.0 \\
\colrule \multirow{2}{*}{$v_2$ (bg)}
    & $+0$ & $+1$ & $+2$ & {~} & {~} & $+2$ & $+4$ & $+2$ \\
    & $-1$ & $-4$ & $-10$ & {~} & {~} & $-10$ & $-1$ & $-1$ \\
\multirow{2}{*}{$v_2$ (n-f)}
    & $+0$ & $+0$ & $+0$ & {~} & {~} & $+0$ & $+0$ & $+0$ \\
    & $-15$ & $-22$ & $-20$ & {~} & {~} & $-15$ & $-22$ & $-20$ \\
\colrule $R_{CP}$ (bg)
    & $\pm4$ & $\pm2$ & $\pm8$ & $\pm2$ & $\pm6$ & $\pm2$ & $\pm4$ & $\pm6$ \\
$R_{CP}$ (eff)
    & $\pm10$ & $\pm10$ & $\pm10$ & $\pm5$ & $\pm9$ & $\pm10$ & $\pm10$ & $\pm10$ \\
\botrule
\end{tabular}
\end{table}

Systematic errors in the calculation of $v_{2}$ are due to
correlations unrelated to the reaction plane (non-flow effects) and
uncertainty in estimates of the background in the invariant mass
distributions. Table~\ref{syserr} lists the dominant systematic
errors.
The systematic error in $v_2$ associated with the yield extraction
(background) is found to be small and the non-flow systematic error is
dominant.
We estimate the non-flow contribution by comparing charged particle
$v_2$ from a reaction-plane analysis and a four-particle cumulant
analysis~\cite{aihong}. The four-particle cumulant analysis is thought
to be insensitive to non-flow effects but leads to larger statistical
errors. Any difference between the methods is assumed to arise from
non-flow contributions. The non-flow contribution to $v_2$ has not
been established experimentally for identified particles.
We examined the effect of standard jet fragmentation on $K_S^0$
and $\Lambda+\overline{\Lambda}$ $v_2$ using superimposed p+p
collisions generated with PYTHIA~\cite{pythia}.
%
Within the measured $p_T$ region, no significant differences are
seen between $\Lambda+\overline{\Lambda}$ and $K_S^0$ non-flow
effects from this source.
We assume a similar magnitude of non-flow contribution to
$\Lambda+\overline{\Lambda}$ and $K_S^0$ $v_2$ and use the difference
between the charged particle $v_2$ from a reaction-plane and a
four-particle cumulant analysis to estimate the upper limit of
possible non-flow contributions to both $\Lambda+\overline{\Lambda}$
and $K_S^0$ $v_2$. Contributions to the systematic errors for $R_{CP}$
come from the determination of the detector efficiency, extraction of
the yields and uncertainty in the model calculation of
$\mathrm{N_{bin}}$~\cite{highpt200}.

\begin{figure}[hbtp]
\centering\mbox{
\includegraphics[width=0.5\textwidth]{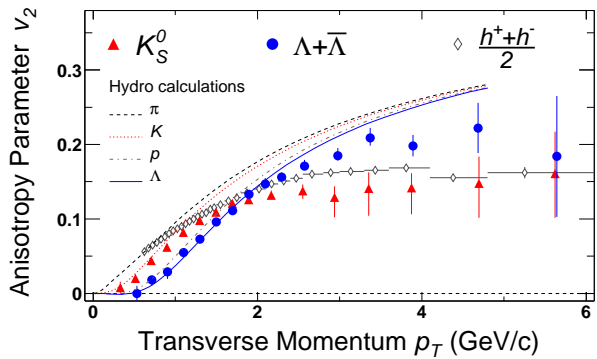}}
\caption{ (color online). The minimum-bias (0--80\% of the collision
  cross section) $v_{2}(p_T)$ for $K_{S}^{0}$,
  $\Lambda+\overline{\Lambda}$ and $h^{\pm}$. The error bars shown
  include statistical and point-to-point systematic uncertainties from the
  background.  The additional non-flow systematic uncertainties are
  approximately -20\%. Hydrodynamical calculations of $v_2$ for pions,
  kaons, protons and lambdas are also
  plotted~\cite{hydroPasi01}.} \label{fig1}
\end{figure}

Fig.~\ref{fig1} shows minimum-bias $v_{2}$ for $K_{S}^{0}$,
$\Lambda+\overline{\Lambda}$ and charged hadrons ($h^{\pm}$). The
analysis method used to obtain the charged hadron $v_2$ is
described in Ref.~\cite{Chgv2}.
Fig.~\ref{fig1} also shows hydrodynamic model calculations of
$v_2$ for pions, kaons, protons, and lambdas~\cite{hydroPasi01}.
At low $p_T$, $v_2$ is consistent with hydrodynamical
calculations, in agreement with the previous results at
$\sqrt{s_{_{NN}}} = 130$~GeV~\cite{v0v2130}.
This Letter establishes the particle-type dependence of the $v_2$
saturation at intermediate $p_T$.
In contrast to hydrodynamical calculations, where at a given $p_T$,
heavier particles have smaller $v_{2}$ values, at intermediate
$p_T$, $v_{2}^{\Lambda} >
v_{2}^{K}$.
The $p_T$ scale where $v_{2}$ deviates from the hydrodynamical
prediction is $\sim 2.5$~GeV/c for $\Lambda+\overline{\Lambda}$
and $\sim 1$~GeV/c for $K_{S}^{0}$.

\begin{figure}[hbtp]
\centering\mbox{
\includegraphics[width=0.5\textwidth]{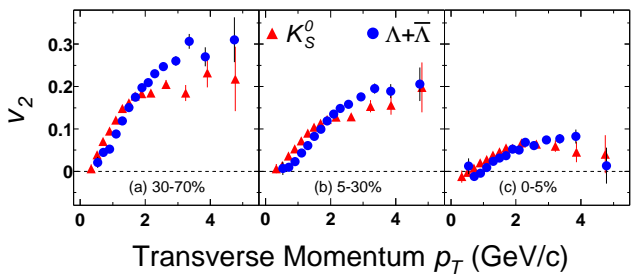}}
\caption{ (color online). The $v_2$ of $K_{S}^{0}$ and
$\Lambda+\overline{\Lambda}$ as a function of $p_T$ for 30--70\%,
5--30\% and 0--5\% of the collision cross section. The error bars
represent statistical errors only. The non-flow systematic errors
for the 30--70\%, 5--30\% and 0--5\% centralities are -25\%, -20\%
and -80\% respectively. } \label{fig2}
\end{figure}


Fig.~\ref{fig2} shows $v_{2}$ of $K_{S}^{0}$ and
$\Lambda+\overline{\Lambda}$ for three centrality intervals:
30--70\%, 5--30\%, and 0--5\%
of the geometrical cross section. 
In each centrality bin, $v_{2}(p_T)$ rises at low $p_T$ and
saturates at intermediate $p_T$.
The values of $v_2$ at saturation are particle-type and
centrality dependent.

\begin{figure}[tbph]
\centering\mbox{
\includegraphics[width=0.5\textwidth]{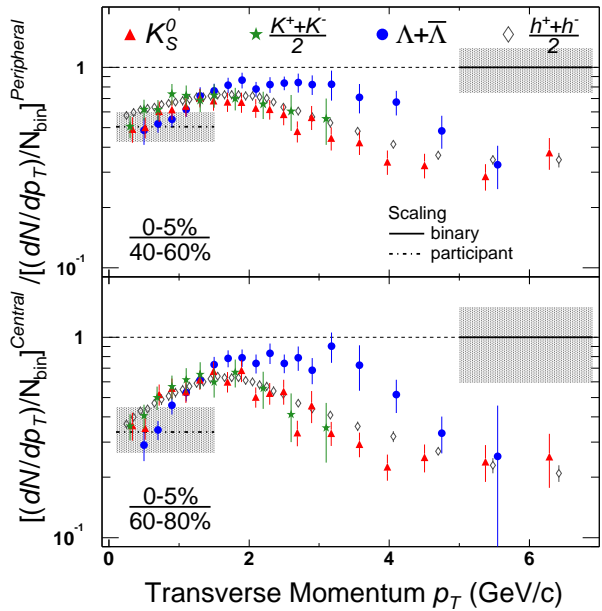}}
\caption{ (color online). The ratio $R_{CP}$ for $K_{S}^{0}$,
  $K^{\pm}$, and $\Lambda +\overline{\Lambda}$ at mid-rapidity
  calculated using centrality intervals, 0--5\% vs. 40--60\% (top) and
  0--5\% vs. 60--80\% of the collision cross section (bottom).  The
  error bars shown on the points include both statistical and
  systematic errors. The widths of the gray bands represent the
  uncertainties in the model calculations of $\mathrm{N_{bin}}$ and
  $\mathrm{N_{part}}$. We also show the charged hadron $R_{CP}$
  measured by STAR for $\sqrt{s_{_{NN}}}=200$~GeV~\cite{highpt200}.}
\label{fig3}
\end{figure}

If partons that fragment into {(anti-)lambdas} lose more energy than
those that fragment into kaons, a particle-type dependence for $v_2$
may develop at high $p_T$ with $v_2^{\Lambda}>v_2^{K}$. In this case,
$\Lambda+\overline{\Lambda}$ yields should be more suppressed than
kaon yields.
Fig.~\ref{fig3} shows $R_{CP}$ for $K_{S}^{0}$, $K^{\pm}$, and
$\Lambda+\overline{\Lambda}$ using the 5\% most central
collisions, normalized by peripheral collisions ($40$-$60\%$ and
$60$-$80\%$).
For charged hadrons, these peripheral bins approximately follow
$\mathrm{N_{bin}}$ scaling without medium
modification~\cite{highpt200}.
The bands in Fig.~\ref{fig3} show the expected values of
$R_{CP}$ for binary and participant ($\mathrm{N_{part}}$) scaling
including systematic variations from the
calculation~\cite{highpt200}.
For most of the intermediate $p_T$ region, $R_{CP}$ for
$\Lambda+\overline{\Lambda}$ is similar to expectations of
$\mathrm{N_{bin}}$ scaling and $R_{CP}^{K} <
R_{CP}^{\Lambda}$.
The $p_T$ scales associated with the saturation and reduction of
$R_{CP}$ also depend on the particle type.
For both species, the $p_T$ where $R_{CP}$ begins to decrease
approximately coincides to the $p_T$ where $v_2$ in
Fig.~\ref{fig1} saturates.
At high $p_T$ ($p_T > 5.0$~GeV/c), $R_{CP}$ values for $K_{S}^{0}$ and
$\Lambda+\overline{\Lambda}$ are consistent with the value for charged
hadron $R_{CP}$, indicating that the baryon enhancement observed at
intermediate $p_T$ in central Au+Au collisions ends at $p_T \approx
5$~GeV/c. The particle-type dependence of $v_2$ and $R_{CP}$ at
intermediate $p_T$ are in contradiction to expectations from energy
loss followed by fragmentation in vacuum.

Nuclear modifications such as shadowing and initial-state
rescattering~\cite{Straub,Accardi} may affect $R_{CP}$ but they are
not expected to give rise to such a large variation with particle-type
(e.g.~\cite{Lev}). At lower beam energy, the enhancement of yields in
p+A collisions at intermediate $p_T$ (\textit{i.e.} the Cronin
effect~\cite{Cronin}) is larger for baryons than
mesons~\cite{Straub}. The Cronin effect has been attributed to
initial-state rescattering, and is expected to decrease with
increasing beam energy~\cite{Accardi}. Alternatively, a strong
particle-type dependence of the Cronin effect may indicate a nuclear
modification to the parton fragmentation. 
Although the effects of shadowing, initial-state rescattering and
non-flow deserve further investigation, the particle-type and $p_T$
dependence of $v_2$ and $R_{CP}$ may reveal a cross-over from a
$p_T$ region dominated by bulk partonic matter hadronization to one
dominated by single parton fragmentation. Our measurements indicate
that the cross-over would occur at $p_T \sim$ 4--5~GeV/c.

\begin{figure}[hbtp]
\centering\mbox{
\includegraphics[width=0.5\textwidth]{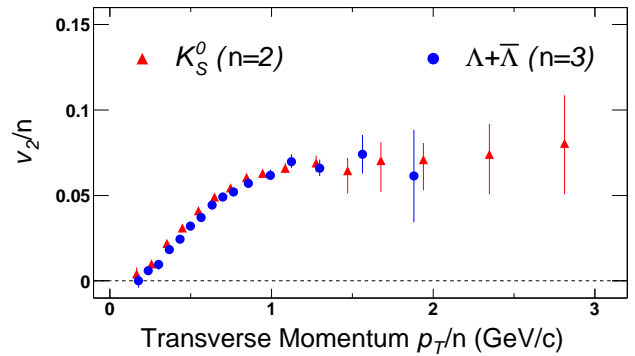}}
\caption{ (color online). The $v_2$ parameter for $K_{S}^{0}$ and
  $\Lambda+\overline{\Lambda}$ scaled by the number of constituent
  quarks (n) and plotted versus $p_{T}/$n. The error bars shown
  include statistical and point-to-point systematic uncertainties from the
  background.  The additional non-flow systematic uncertainties are
  approximately -20\%.} \label{fig4}
\end{figure}

The larger $\Lambda+\overline{\Lambda}$ $R_{CP}$ at intermediate
$p_T$ shows that the $\Lambda+\overline{\Lambda}$ yield
increases with parton density faster than the kaon yield.
Multi-parton mechanisms such as gluon junctions~\cite{Vance}, quark
coalescence \cite{CoalVoloshinv2}, or recombination
\cite{CoalMullerRaa} can naturally lead to a stronger dependence on
parton density for baryon production than meson production.
Models using coalescence or recombination mechanisms in particle
production predict that at intermediate $p_T$ $v_2$ will follow a
number-of-constituent-quark scaling~\cite{CoalVoloshinv2}.
Fig.~\ref{fig4} shows $v_2$ of $K_{S}^{0}$ and
$\Lambda+\overline{\Lambda}$ as a function of $p_T$, where the
$v_2$ and $p_T$ values have been scaled by the number of
constituent quarks (n).
While $v_2$ is significantly different for $K_S^0$ and
$\Lambda+\overline{\Lambda}$, within errors, $v_2$/n vs $p_T$/n is
the same for both species above $p_T$/n $\sim 0.7$~GeV/c.
In a scenario where hadrons at intermediate $p_T$ coalesce from
co-moving quarks, $v_2/$n($p_T/$n) reveals the momentum-space
azimuthal anisotropy of partons in a bulk matter~\cite{CoalVoloshinv2}.

At higher $p_T$ where independent fragmentation is likely to dominate
over multi-parton particle production mechanisms,
constituent-quark-scaling is expected to break down and the $K_S^0$
and $\Lambda+\overline{\Lambda}$ $v_2$ may take on a value closer to
that of an underlying partonic $v_2$~\cite{CoalVoloshinv2}. The
convergence of $K_S^0$ and $\Lambda+\overline{\Lambda}$ $R_{CP}$ at
$p_T \sim 5$~GeV/c in Fig.~\ref{fig3} supports this
expectation. Higher statistics $v_2$ measurements in this region along
with measurements of $v_2$ for other identified particles will
therefore provide an opportunity to test the scaling demonstrated in
Fig.~\ref{fig4}.

In summary, we have reported the measurement of $v_2$ and $R_{CP}$
up to $p_T \sim 6.0$~GeV/c for kaons and
$\Lambda+\overline{\Lambda}$ from Au+Au collisions at
$\sqrt{s_{_{NN}}}=200$~GeV.
At low $p_T$, hydrodynamic model calculations agree well with
$v_2$ for $K_S^0$ and $\Lambda+\overline{\Lambda}$.
At intermediate $p_T$, however, hydrodynamics no longer describes
the particle production.
For $K_{S}^{0}$, $v_2$ saturates earlier and at a lower value than
for $\Lambda+\overline{\Lambda}$.
The $K_{S}^{0}$ and $\Lambda+\overline{\Lambda}$ $v_2$ are shown
to follow a number-of-constituent-quark scaling law.
In addition, $R_{CP}$ shows that the yield of
$\Lambda+\overline{\Lambda}$ is increasing more rapidily with the
system size than kaons: At intermediate $p_T$, the
$\Lambda+\overline{\Lambda}$ $R_{CP}$ is close to expectations
from binary scaling while the kaon $R_{CP}$ is lower.
At high $p_T$, the $R_{CP}$ of $K_{S}^{0}$ and
$\Lambda+\overline{\Lambda}$ are consistent with the value for charged
hadrons, indicating that the centrality dependent baryon enhancement
observed at intermediate $p_T$ ends near $p_T = 5$~GeV/c.
The measured features at intermediate $p_T$ are consistent with the
presence of multi-parton particle formation mechanisms beyond the
framework of parton energy loss followed by standard fragmentation.
The particle- and $p_T$-dependence of $v_2$ and $R_{CP}$
constitute a unique means to investigate the anisotropy and
hadronization mechanism of the bulk dense matter formed in
nucleus-nucleus collisions at RHIC.

\vspace{0.25cm} \textbf{Acknowledgments:} We thank the RHIC
Operations Group and RCF at BNL, and the NERSC Center at LBNL for
their support. This work was supported in part by the HENP
Divisions of the Office of Science of the U.S. DOE; the U.S. NSF;
the BMBF of Germany; IN2P3, RA, RPL, and EMN of France; EPSRC of
the United Kingdom; FAPESP of Brazil; the Russian Ministry of
Science and Technology; the Ministry of Education and the NNSFC of
China; SFOM of the Czech Republic, DAE, DST, and CSIR of the
Government of India; the Swiss NSF.

\end{document}